\documentclass{jpsj2}
\title{Quantum Kagome antiferromagnet ZnCu$_3$(OH)$_6$Cl$_2$.}

\author{\textsc{Philippe Mendels}$^{1}$\thanks{E-mail: mendels@lps.u-psud.fr} and \textsc{Fabrice Bert}$^{1}$}

\inst{$^{1}$Laboratoire de Physique des Solides, Universit\'e Paris-Sud XI Orsay, UMR 8502 CNRS, 91405 Orsay Cedex, France \\}

\abst{The frustration of antiferromagnetic interactions on the loosely connected kagome lattice associated to the enhancement of quantum fluctuations for S=1/2 spins was acknowledged long ago as a keypoint to stabilize novel ground states of magnetic matter. Only very
recently, the model compound Herbersmithite, ZnCu$_{3}$(OH)$_6$Cl$_2$, a structurally perfect kagome antiferromagnet, could be synthesized and enables a close comparison to theories. We review and classify various experimental results obtained over the past years and
underline some of the pending issues.}

\kword{frustration, kagome, spin liquids, quantum magnetism}

\begin{document}
\maketitle

\section{Introduction}

After the initial papers by Anderson and Fazekas in 1973-74~\cite{Fazekas} and the revival of their seminal mainstream proposal of a resonating valence bond ground state in the context of High Temperature Superconductors~\cite{Anderson}, kagome antiferromagnets have been recognized since the early 90's as the corner stone of the search for a quantum spin liquid state. Strong indications for very original physics had been found in some experimental systems with a kagome based geometry, such as magnetic freezing occuring only at very low temperatures on the scale of the exchange energy and fluctuating ground states, two features recognized as the landmarks of frustration~\cite{Ramirez90,Uemura}. Only recently, a "New Candidate emerged for a quantum spin liquid"~\cite{Levy}. Namely, Herbertsmithite, a rare mineral~\cite{Braithwaite} which has been synthesized for the first time in 2005~\cite{Shores} has brought a tremendous excitement to the field. Altogether with $\kappa$(ET)$_2$Cu$_2$(CN)$_3$, it combines low dimensionality, a quantum character associated with spins 1/2 and the absence of magnetic freezing.  This was celebrated as "An End to the Drought of Quantum Spin  Liquids"~\cite{PLee} in the context of a grail quest for the synthesis of a kagome model system to compare with theories.
Although theoretical models have been developed over the last 20 years, there is still no consensus about the expected ground state. Most of the models point at an exotic ground state and two broad classes of spin liquids have been explored theoretically, in particular for the kagome antiferromagnet, (i) topological spin liquids which feature a spin gap, a topological order and fractional excitations~\cite{theory1}; (ii) algebraic spin liquids with gapless excitations in all spin sectors~\cite{theory2}.

Here, we focus on the review of the experimental properties of Herbertsmithite and discuss the deviations to the ideal case and their implications with regards to the important issue of the existence of a gap.

\section{Structure, interactions and phase diagram of the generic family of paratacamites, Zn$_{x}$Cu$_{4-x}$(OH)$_6$Cl$_2$, $0<x<1$}

\begin{figure}[h]
\begin{center}
\includegraphics*[width=1\textwidth]{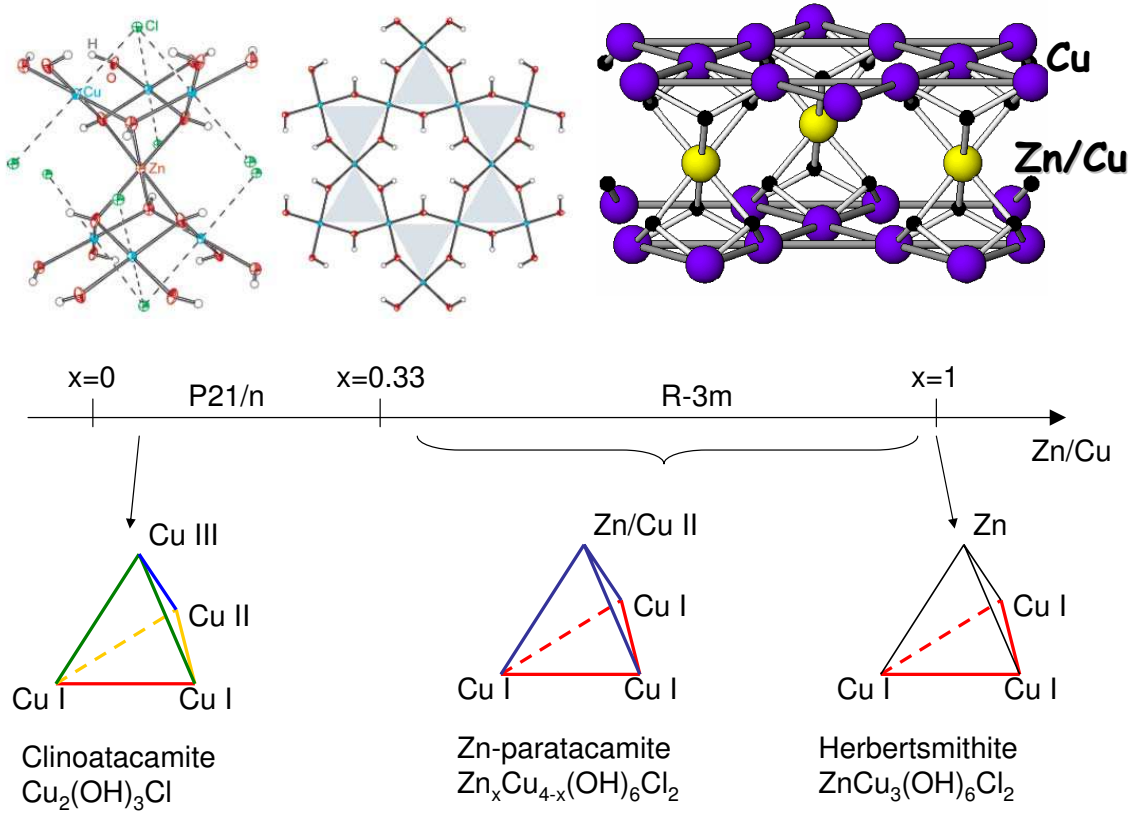}
\end{center}
\caption{(Color online) Top left: Structure of Herbertsmithite from~\cite{Shores}. Top right: Simple sketch of the structure where Zn and Cu sites only have been represented. Bottom: Evolution of the structure of paratacamites when Zn content is increased. A structural transition occurs around $x=0.33$ which yields well defined perfect kagome planes, assumed to be filled with Cu only.}
\label{structure_Herbert}
\end{figure}

Herbertsmithite ZnCu$_3$(OH)$_6$Cl$_2$ is the $x=1$ end compound of the Zn-paratacamite family Zn$_x$Cu$_{4-x}$(OH)$_6$Cl$_2$~\cite{Braithwaite,Shores}. It can be viewed as a double variant of the parent clinoatacamite compound ($x=0$), situated at the other end of the family. Starting from the latter, a Jahn-Teller distorted $S=1/2$ pyrochlore, the symmetry first relaxes from monoclinic ($P2_1/n$) to rhombohedral ($R\overline{3}m$) around $x=0.33$, leading to a perfect kagome lattice in the $a$-$b$ plane with isotropic planar interactions; then, in the $c$- elongated $x>0.33$ pyrochlore structure, the magnetic bridge along $c$-axis between $a$-$b$ kagome planes is progressively suppressed by replacing the apical Cu$^{2+}$ by a diamagnetic Zn$^{2+}$. Due to a more favorable electrostatic environment, Cu$^{2+}$ is expected to preferentially occupy the distorted octahedral kagome sites only. When $x=1$ the $S=1/2$  ions should therefore form structurally perfect kagome layers that are themselves well separated by diamagnetic Zn$^{2+}$(Fig.~\ref{structure_Herbert}).

\subsection{Clinoatacamite}

In clinoatacamite ($x=0$) two successive transitions to long-range ordered phases are observed at 18~K and 6~K. They are signalled by the development of well defined internal fields in $\mu$SR~\cite{Zheng-b}, by peaks in specific heat and accidents in magnetic susceptibility~\cite{Zheng-a}. At the 6~K transition a ferromagnetic component along the $b$-axis develops which likely relates to some Dzyaloshinskii-Moriya anisotropy and/or weak ferromagnetic interactions between the planes~\cite{Kim, clinoatacamite_Wills2}. While in the low-$T$ phase a combination of three planar interactions, as expected from the low symmetry of the structure, ($\overline{J}_{planar}\sim 53$~K) and anisotropy $\sim 0.08$~$\overline{J}$ are necessary to explain the inelastic neutron data~\cite{Kim}, there are still inconsistencies between the results from different techniques. On one hand, $\mu$SR indicates that the order parameter fully develops at the upper transition (18~K). Below 6~K, besides some rearrangement of the magnetic structure which might explain the change in the internal field at the muon sites, an unexpected enhancement of fluctuations is observed~\cite{Zheng-b}. On the other hand, only a weak amount of entropy is released at 18~K and a clear signature of long-range ordering appears in neutron diffraction only below 6~K, suggesting a highly frustrated state with weak moments between 6 and 18~K in strong contradiction with $\mu$SR results~\cite{Zheng-a, clinoatacamite_Wills2}.

\subsection{Phase diagram}
\begin{figure}[h]
\begin{center}
\includegraphics*[width=28pc]{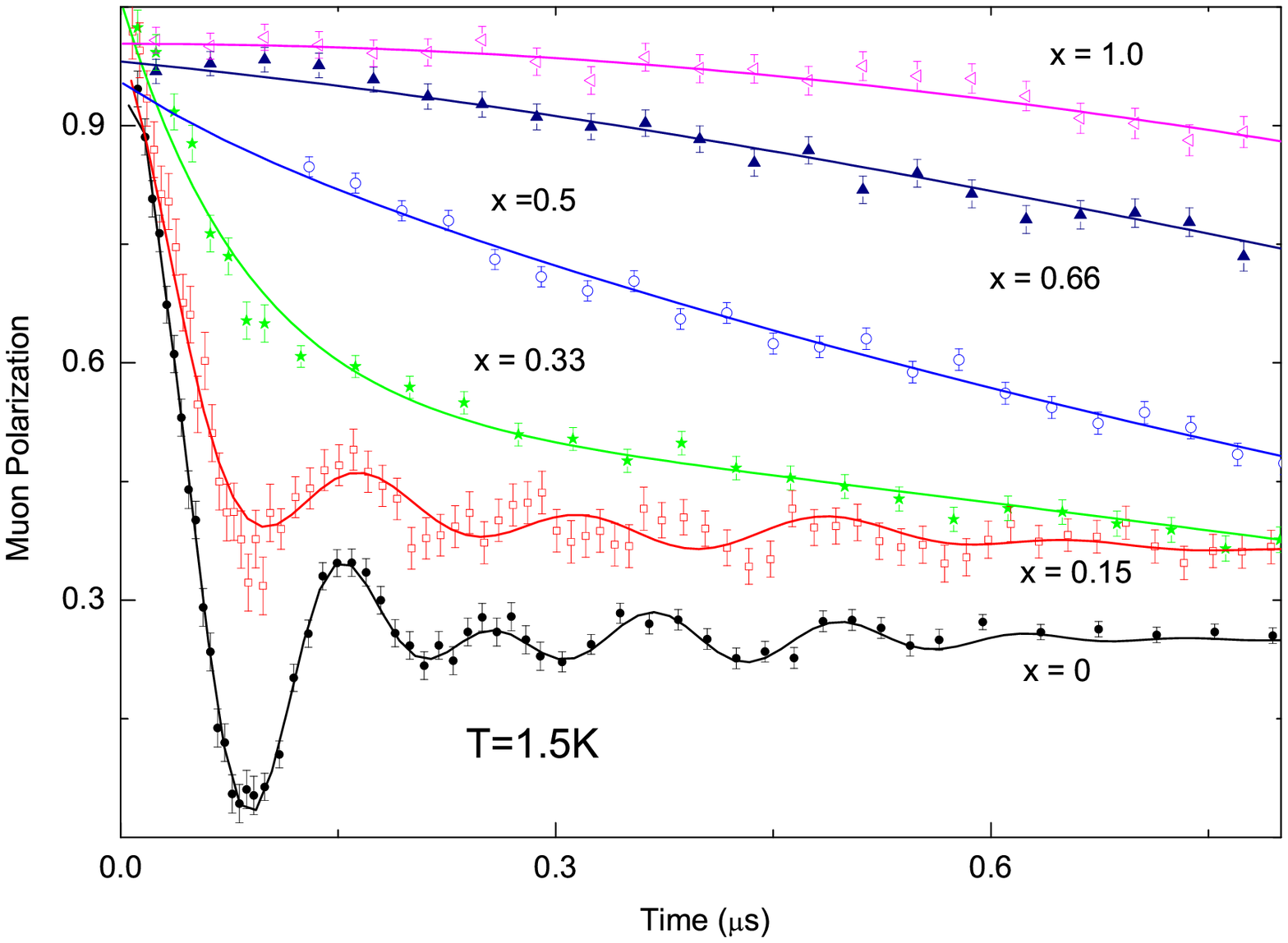}
\end{center}
\caption{(Color online) Evolution of the low-$T$ $\mu$SR asymmetry when Zn content is progressively increased. The oscillations due to the existence of well defined static fields progressively disappear and, for $x>0.5$, the frozen magnetism disappears. (Adapted from~\cite{MendelsHerbert})}
\label{muonpolarization}
\end{figure}
\begin{figure}[tb]
\centering
\includegraphics*[width=22pc]{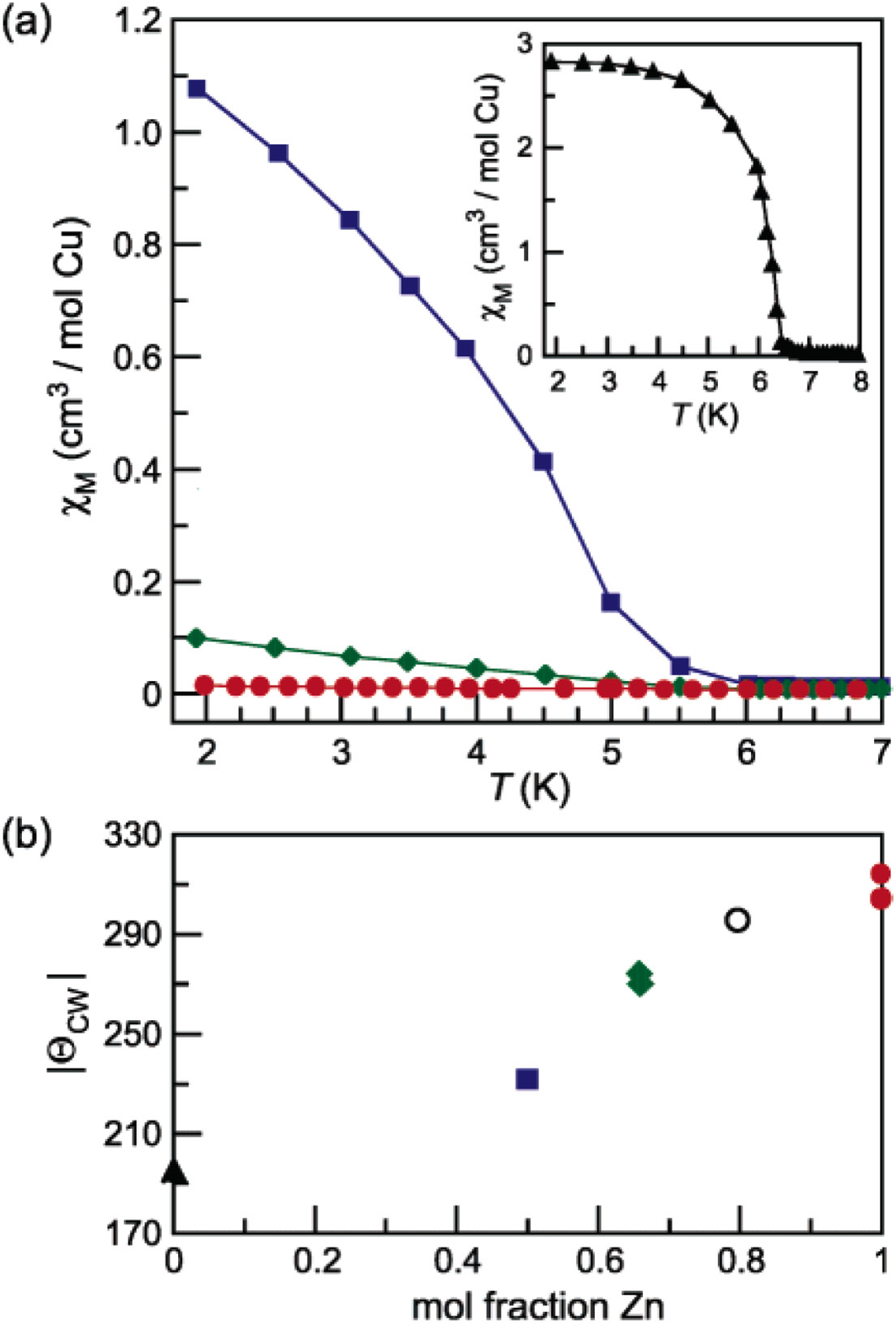}
\caption[]{(Color online) Top panel: Susceptibility versus $T$ for $x=0$, 0.5, 0.66 and 1.0 in the paratacamite family. The corresponding symbols can be read from the bottom panel. The $T=0$ ferromagnetic component decreases progressively from $x=0$ (inset) to $x=1.0$. Bottom panel: The apparent Curie-Weiss temperatures are extracted from linear fits of the inverse susceptibility up to 300~K. These values have to be corrected in order to extract the exact value of the exchange interaction, (see text). Adapted from~\cite{Shores}.}
\label{chi_Shores}
\end{figure}
\begin{figure}[h]
\centering
\includegraphics*[width=28pc]{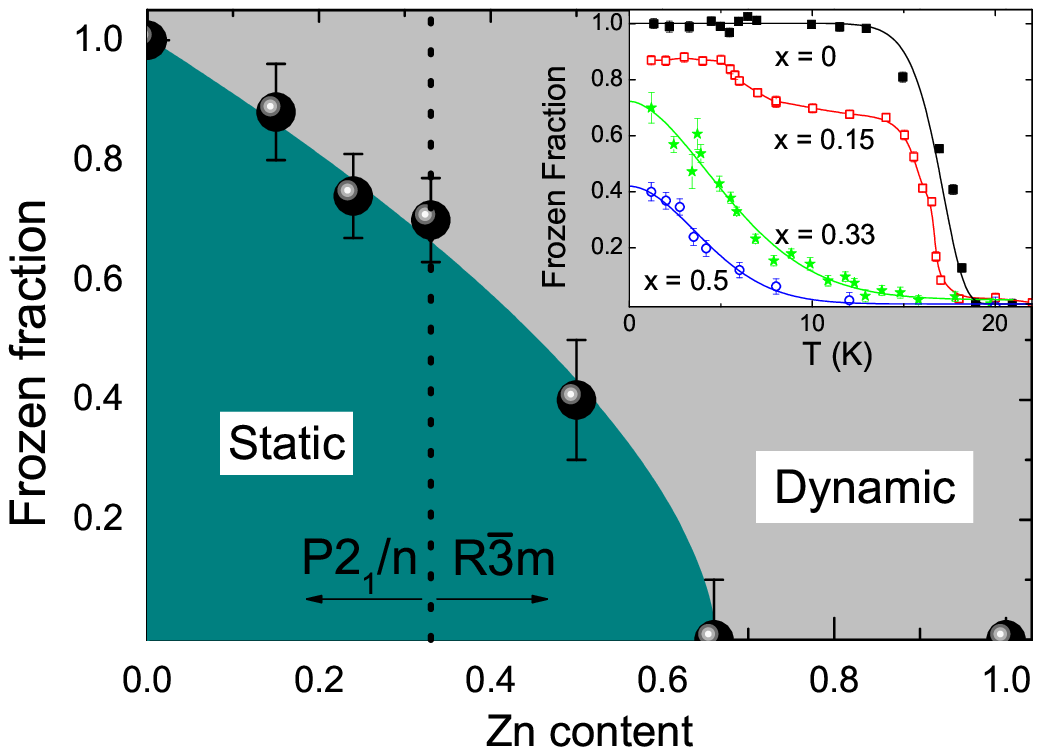}
\caption[]{(Color online) Phase diagram of the paratacamite family as deduced from the $T\rightarrow 0$ frozen fractions found in zero field $\mu$SR experiments~\cite{MendelsHerbert} (inset).}
\label{muons Herbert}
\end{figure}

Owing to its extreme sensitivity to static frozen phases, zero field $\mu$SR experiments enable to track in minute details the magnetism of the entire paratacamite family~\cite{MendelsHerbert}. For intermediate Zn contents $0.33<x<0.66$, the low temperature phase appears to be locally inhomogeneous and yields two components in the $\mu$SR signal; one from a frozen magnetic environment reminiscent of the ordered phase in clinoatacamite which gets more and more disordered as $x$ increases, and one from a dynamical environment of the muon (Fig.\ref{muonpolarization}). The evolution of this frozen fraction detected in $\mu$SR yields the phase diagram plotted in Fig.~\ref{muons Herbert}. Interestingly for $x>0.66$, the frozen fraction gets lower than the experimental accuracy (a few percent) and a fully "liquid" phase appears although 1/3$^{rd}$ of the inter-layer 3D coupling paths are still active. Accordingly, the small ferromagnetic component found in macroscopic magnetization measurements disappears progressively when going from $x=0$ to 1 (Fig.\ref{chi_Shores}).

The large domain of stability of a "liquid" phase in the phase diagram of paratacamites likely points at the weakness of the interplane coupling as well as at the shortness of the in plane correlation length.
No magnetic freezing has ever been observed in the $x=1$ compound  down to the lowest probed temperature of 20~mK~\cite{MendelsHerbert}.
The upper bound value of a hypothetical frozen moment would then be less than 6 10$^{-4} \mu_{\rm B}$ ($\mu$SR). This is completely consistent with ac susceptibility results for $x=1$,  which behaves monotonously down to the lowest temperature (50~mK) and with neutron scattering where no magnetic diffraction Bragg peak was detected at variance with clinoatacamite~\cite{Helton}.

Herbertsmithite therefore appears as the first structurally perfect kagome antiferromagnet not displaying any magnetic transition and opens wide the field of experimental investigations in quantum kagome antiferromagnets.

\subsection{Interactions}

In Herbertsmithite, the $119 ^{\circ}$ Cu-OH-Cu bond-angle yields a moderate planar interaction $J\simeq 180(10)$~K in comparison with other cuprates. Note that this value is consistent with that observed in cubanes where a similar OH bridge links adjacent coppers~\cite{clinoatacamite_Wills2}. 
From a simple Curie-Weiss analysis of the variation of the high-$T$ susceptibility with $x$, the interaction between apical and planar Cu can be estimated to be weakly ferromagnetic, $J^{'} \sim 0.1 J$, in agreement with neutron low-$T$ data on the $x=0$ compound. The paratacamite family therefore offers an interesting possibility to explore weakly ferromagnetically coupled kagome layers~\cite{Shores} for $0.33<x<1$.

\section{Magnetism of the kagome planes}

\subsection{Static susceptibility}

Probing the susceptibility is a crucial issue regarding the comparison and discrimination between models. It has represented a major challenge since the discovery of the compound. Various results are now reported through macroscopic susceptibility, and local techniques, by shift NMR measurements through $^{35}$Cl~\cite{Imai} and $^{17}$O NMR~\cite{Olariu} or $\mu$SR~\cite{Ofer}. Among all these local probes, $^{17}$O NMR is certainly the most sensitive to the physics of the kagome planes since oxygen from the OH group mediates the superexchange coupling between two adjacent Cu. The hyperfine coupling of Cl to Cu is found more than one order of magnitude smaller, itself much larger than the muon one. We will focus first on $^{17}$O NMR.

Typical $^{17}$O NMR spectra are displayed in Fig.~\ref{spectraO17}. From a detailed analysis of the marked singularities of the powder lineshape which positions are governed by the magnetic shift and quadrupolar effects, two different sites can be identified. The most intense line at high $T$, named main(M), is associated with O sites linked to two Cu sites in the kagome planes. Tracking its position gives the shift, hence one can extract the hereafter called "intrinsic" susceptibility  from 300~K down to 0.45~K. The other line (D) which is very sharp at low-$T$ is characteristic of O sites linked with one Cu and one Zn in the kagome planes. The existence of this line points at the existence of defects resulting mainly from Cu-Zn site disorder. This will be further discussed in the next sections as well as the corresponding details of the $^{17}$O NMR spectra such as the broadening observed at low $T$ for the (M) line.

Cl NMR shift could be as well followed on partially oriented powders but due to a large broadening when $T$ is decreased, it becomes impossible to determine the susceptibility below 15~K.

It is worth noting that no quantitative derivation of the couplings of the probes to the various Cu sites has been made up to now but the diversity of the observed behaviors at low $T$ might likely be associated to the ratios of couplings strength of each probe to the various Cu locations including the non-kagome Zn site~\cite{footnote}.

\begin{figure}[t]
\begin{center}
\includegraphics[width=28pc]{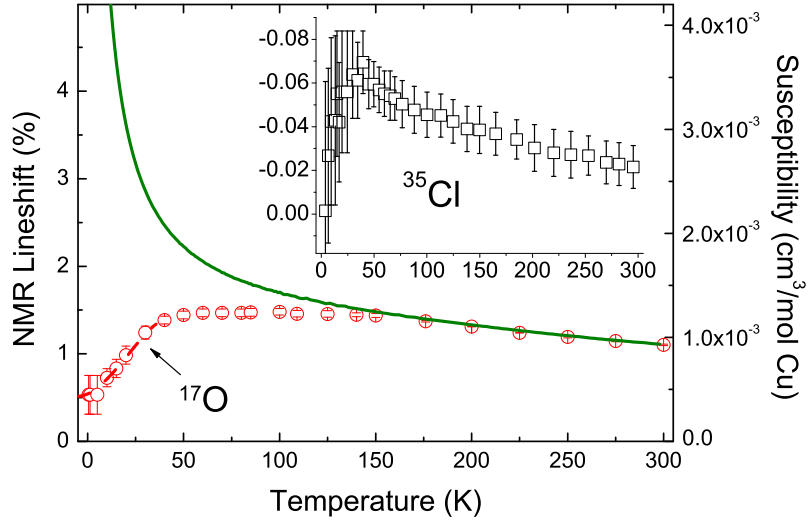}
\caption{(Color online) Thermal variation of the susceptibility of
the kagome planes measured from the shift of the main $^{17}$O NMR
line (red open symbols) compared to the macroscopic susceptibility
( green solid line). The red dashed line is a guide to the eye for the low-$T$ shift. Inset: $^{35}$Cl data show the same trend as $^{17}$O data although the accuracy is much reduced below 50~K. Adapted from ref.~\cite{Olariu, Imai}.}
\label{fig_shift}
\end{center}
\end{figure}

One can divide the behavior of the susceptibility into 3 $T$-ranges (Fig.~\ref{fig_shift}):
\begin{figure}[t]
\centering
\includegraphics*[width=0.5\textwidth]{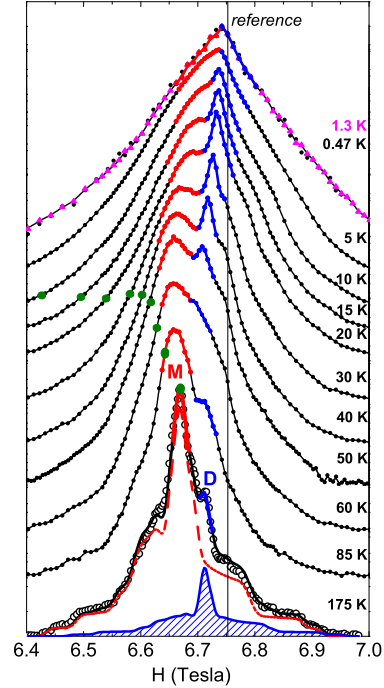}
\caption[]{(Color online) $^{17}$O NMR spectra from 175~K to 0.47~K from ref.~\cite{Olariu}. The shift reference is indicated by the vertical line. (M) and (D) refer to the two oxygen sites in the kagome planes as described in the text. Large full green circles indicate the position of the center of the line that would be expected from $\chi_{macro}$. At 175~K, a detailed analysis was performed and two contributions are evident from the displayed fits (see text and section 4). The hatched area represents the contribution from (D) oxygens sitting nearby one Zn defect in the kagome plane.}
\label{spectraO17}
\end{figure}

\subsubsection{High $T$: $T>150$~K}

In macroscopic susceptibility ($\chi_{macro}$) experiments as well as for $^{17}$O NMR, or $^{35}$Cl NMR, an effective Curie-Weiss behavior is observed at high $T$. From a correct fit using high temperature series expansions, one can extract a value $J=170(10)$~K from ~\cite{RigolSingh1, MisguichHerbert}.

\subsubsection{Intermediate $T$: $10<T<150$~K}

Below 150~K, macroscopic and local susceptibilities as measured from $^{35}$Cl and $^{17}$O NMR, start to markedly differ as seen directly from Fig.~\ref{spectraO17} and \ref{fig_shift}. A clear upturn is observed in $\chi_{macro}$ whereas the variation of the oxygen shift goes the opposite direction. $^{35}$Cl NMR data also show a deviation from $\chi_{macro}$. This is certainly the best evidence that the compound is not free from defects. Would it be, the only contribution to $^{17}$O and $\chi_{macro}$ measurements would come from fully occupied Cu$^{2+}$ kagome planes and would be identical provided that no staggered component dominates the O signal. Through $^{17}$O shift measured from 300~K $\sim 1.7$ $J$ down to 0.45~K $\sim J/400$, a broad crossover with a smooth maximum from the Curie-Weiss to the low-$T$ regime is observed between $\sim 2J/3\sim$ 120~K and $\sim J/3\sim$ 60~K, see Fig.~\ref{fig_shift}. Such maxima are also observed in other corner-sharing kagome based lattices~\cite{Mendels_SCGO,Limot,Bono1,Hiroi} but the $T$ dependence of the shift differs from one compound to another. A rapid decrease of the shift is further observed at lower $T$, between $J/3$ and $J/20\sim 10$~K.

\subsubsection{Low $T$: $T < 5-10$~K}
For $T < 10$~K$\sim J/20$, the shift finally levels off at a \emph{finite} value. For instance, no difference can be found between the $^{17}$O spectra taken at 0.45 and 1.3~K. The non-zero value and the $T$ dependence of the shift lead to the  conclusion that the kagome plane susceptibility of Herbertsmithite  has a \emph{finite} value at $T\rightarrow 0$ and the system is probably \emph{gapless} .

\subsection{Dynamical susceptibility}
 \begin{figure}[h]
\begin{center}
\includegraphics[width=25pc]{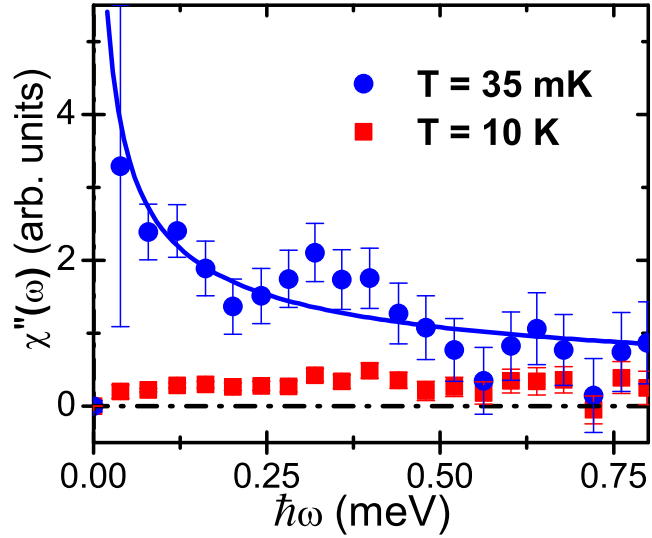}
\caption{(Color on line) Imaginary part of the dynamical susceptibility integrated over momentum transfer. The apparently divergent behavior is indicated by a power law fit, as represented by the full line on the 35~mK data. The exponent -0.5(3) is in line with the -0.7(3) obtained from earlier data, in ref.~\cite{Helton}.}
\label{dynamic-susc}
\end{center}
\end{figure}
On the dynamical side, inelastic neutron scattering (INS) data taken at 35~mK (Fig.~\ref{dynamic-susc}), indicate an increase of $\chi''(\omega)$ with decreasing $\omega$ and clearly supports the absence of a spin gap at least larger than 0.1~meV. The apparently divergent power law behavior $\chi''(\omega)\sim \omega^{-0.7(3)}$ is also quite unusual. Moreover $\chi''$ is only weakly $Q$ dependent and points at the absence of a characteristic length to describe the correlations~\cite{Helton}.

\begin{figure}[h]
\centering
\includegraphics*[width=28pc]{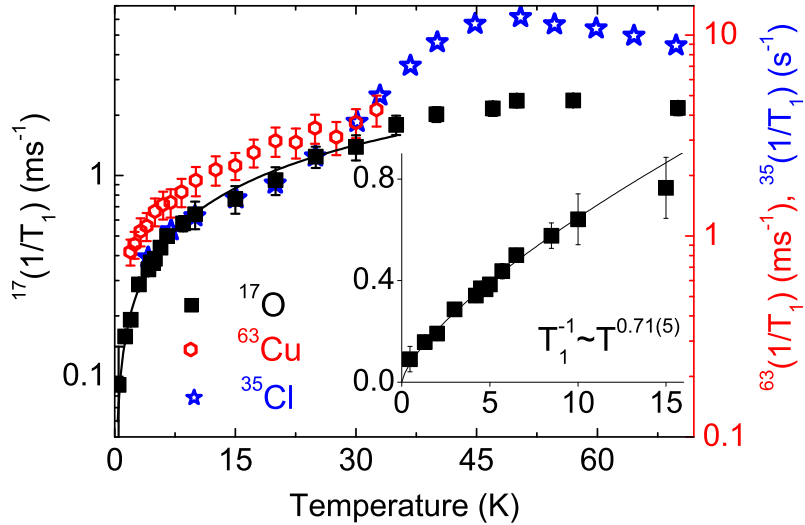}
\caption[]{(Color on line) Semi-log plot of the relaxation rates obtained for all probe nuclei, $^{17}$O, $^{63}$Cu and $^{35}$Cl. Inset: Linear plot of $T_1^{-1}$ with a power law fit for the low $T$ $^{17}$O data. Adapted from ref.~\cite{Olariu, Imai}.}
\label{T1}
\end{figure}

$T_1$ relaxation measurements complement the insight into the structure of the excitation spectrum. The thermal dependence of the relaxation rate obtained for $^{17}$O~\cite{Olariu}, $^{35}$Cl and $^{63}$Cu \cite{Imai} NMR are plotted in Fig.~\ref{T1}.  Quite strikingly, in the low $T$ regime ($T<30$~K$\sim J/6$), the high field relaxation rate of O, Cl and Cu nuclei follow the same rather unusual variation, namely a power law with the sublinear exponent $T_1^{-1}\sim T^{0.71(5)}$~\cite{Olariu, Imai}. Given the filtering form factors which are quite different for all probes, this indicates that excitations are likely weakly $Q$-dependent. The $T$-dependence clearly demonstrates the absence of a gap in the excitations and confirm the neutron results. Note that a very different $T$-behavior, namely a moderate yet observable increase of the relaxation followed by a saturation is observed in $\mu$SR experiments for $T<1$~K but this might be related to weak interactions between magnetic defects~\cite{MendelsHerbert}.

\section{Magnetic defects and their contribution to macroscopic measurements}

In addition to the previously reported discrepancy between $^{17}$O shift and macroscopic susceptibility, there are several types of arguments which consistently fit with a scenario of dilution of the kagome magnetic lattice of Herbertsmithite. This comes out from inter-site mixing defects of Cu and Zn species even for compounds with the ideal stoechiometry.
Namely, part of the non-magnetic Zn$^{2+}$ substitutes copper on the kagome plane, and respectively, to preserve stoichiometry, the same amount of copper substitutes zinc on the inter-plane site. One should note that samples of different origins yield very similar results in macroscopic measurements, $^{35}$Cl NMR and $\mu$SR, so that the various interpretations which have been successively proposed relate rather with progress in the understanding of Herbertsmithite physics than with differences between samples.

It is fair to point that a different picture of perfectly filled Cu$^{2+}$ kagome planes was initially favored based on a quite funded Jahn-Teller argument~\cite{Shores}. Indeed, since Cl$^{-}$ are quite far from Cu$^{2+}$ located in the kagome planes, the quasi-planar configuration is thought to be most favorable to Cu occupation through Jahn-Teller stabilization. This statement was reinforced by the interpretation of early shift measurements through $\mu$SR which was not then confirmed by other local probes. Yet, as pointed above, it is today quite hard to explain all the experimental findings within the initial framework  of what was initially hailed as a "perfect" compound. The Jahn-Teller energy stabilization might not be large enough to completely prevent inter-site mixing and it would be interesting to know more about the Cu Jahn-Teller energy in this compound.

\subsection{Direct evidence of intersite mixing defects}

Although Zn and Cu have close neutron scattering cross-sections, structural refinements are sensitive enough, as compared to x-ray ones, to suggest a large amount 7-10$\%$ of such inter-site mixing defects~\cite{deVries,Lee07}, depending on the origin of the samples.

\begin{figure}[h]
\centering
\includegraphics*[width=28pc]{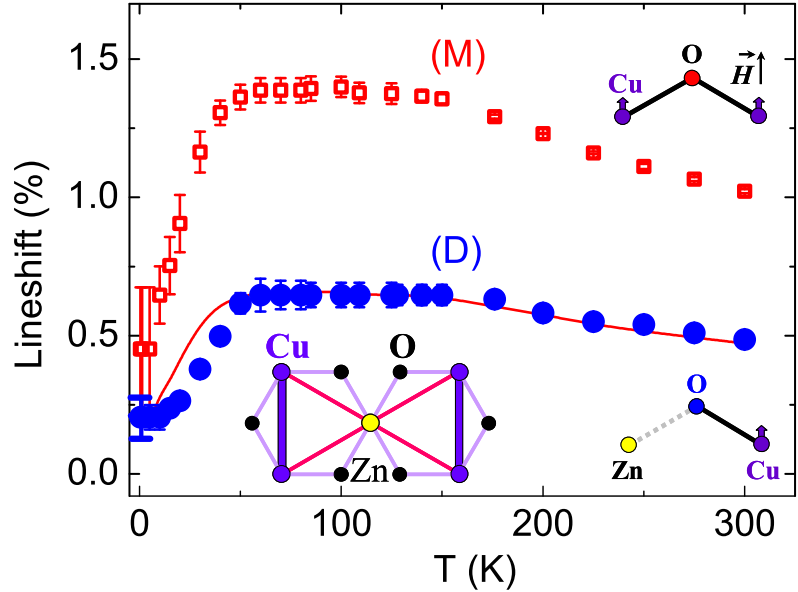}
\caption[]{(Color on line) Comparison between the shift of the main line (M: open red squares) and that from the defect line(D: full blue circles (adapted from ref.~\cite{Olariu}). The corresponding oxygen environment are displayed at the right of the figure. The red line represents half the M shift which is a correct approximation of the D shift down to $J/4$. Lower in temperature, the D shift is found to decrease more rapidly. This indicates a depressed susceptibility near a spin vacancy. The sketch at the bottom left of the figure represents what is expected in an ideal case where singlet dimers are formed next to a spin vacancy~\cite{Dommange03}. Dzyaloshinskii Moriya anisotropy slightly corrects this singlet picture~\cite{Rousochatzakis}.}
\label{Shift-defect}
\end{figure}

NMR spectra taken at high $T$, in the uncorrelated regime, see Fig.~\ref{spectraO17}, give a local evidence for this. Indeed, the NMR powder average spectrum displays many singularities which correspond to singular orientations of shift and quadrupolar tensors. Although oxygen occupies only one crystallographic site in Herbertsmithite, it was found impossible to fit such a powder lineshape~\cite{Bert09} with only one O site. The additional site is well explained by oxygens surrounded by one Cu$^{2+}$ contributing to its shift and one Zn$^{2+}$, ie a spin vacancy which does not contribute to the shift as sketched in Fig.~\ref{Shift-defect}. A quantitative estimate of a 5\% magnetic dilution was calculated from the ratio of intensities of this "defect" (D) line to that of the main (M) line. The shift of this second site is reported in Fig.~\ref{Shift-defect} and compared to half the shift of the main line. These two quantities match well in the high-$T$ regime where correlations are negligible but differences occur at low-$T$. In addition, the relative sharpness of the defect-line clearly point at some different low-$T$ physics for the defect sites. We postpone the related discussion to the next section but, interestingly, the NMR lineshape can be satisfyingly reproduced in exact diagonalization of small spin clusters on the kagome lattice with random spin vacancies~\cite{Chitra08RMN}.

Finally, the low energy magnetic spectral weight in inelastic neutron scattering is found to shift linearly with field towards higher energies, which is commonly interpreted as a Zeeman splitting of paramagnetic-like $S=1/2$ centers. Yet, because of linewidth and scattering intensity arguments developed in~\cite{Helton, Lee07} they cannot be interpreted as arising from strictly non-interacting spins. Either weakly coupled coppers occupying the Zn site due to intersite mixing, or paramagnetic centers  due to spinless defects (see below) could yield a qualitative interpretation for this.

\subsection{Macroscopic measurements: quasi-free Cu$^{2+}$ spins on the Zn$^{2+}$ site}

\subsubsection{Susceptibility}

The intersite mixing is responsible for two kinds of magnetic defects that can show up in the susceptibility of Herbertsmithite.

First, the misplaced Zn$^{2+}$ in the kagome plane act as spinless impurities which are likely to induce a staggered
magnetization of the nearby Cu$^{2+}$ spins as commonly observed in strongly correlated systems~\cite{alloul07RMP}. This could well relate to the magnetic broadening of the $^{17}$O line at low $T$, see Fig.\ref{spectraO17}. Probing the response to such defects is a "perturb to reveal" strategy which might prove to be rewarding. A special section is devoted to this in the discussion part.  As observed in the kagome-like S=1/2 Volborthite compound, where non-magnetic impurities can be introduced in a controlled manner, one can then expect that the whole extended defect around the spin vacancy will behave as a paramagnetic center with a small effective moment~\cite{Bert04}.  The corresponding contribution is often omitted in the context of Herbertsmithite and ought to be considered.

Second, the misplaced Cu$^{2+}$ on the Zn interlayer site are weakly coupled to the kagome planes if one refers to the weak interkagome plane coupling in the paratacamite family (see section 2.3), and can be considered as quasi-free S=1/2 spins. While the first type of defects is well studied by $^{17}$O NMR, the second type is a classical case of weakly coupled paramagnetic centers, which give the dominant contribution, at low temperature, to the macroscopic susceptibility. This is indeed what is observed in these macroscopic measurements. From a simple Curie analysis of the low-$T$ Curie-like tail of the susceptibility, one gets an estimate of 5-10\% of such Zn-defects in the kagome planes.

\subsubsection{Low $T$ and High Field Magnetization}

In low $T$ magnetization measurements under strong magnetic fields (see Fig.~\ref{fig_magnetization})~\cite{BertHerbert}, the magnetization is linear at low field and shows a pronounced downward curvature above $\sim 2$~T as expected for the Brillouin type saturation of free spins, although a saturation plateau at higher field is not observed. Instead, the magnetization for $H \gtrsim 8$~T increases rather linearly with the applied field. As a first approximation, one can split the total magnetization as the sum of two terms.

First, a linear term that is likely to arise from the susceptibility of the Cu$^{2+}$ spins of the kagome planes. Due to their strong
antiferromagnetic interactions, their magnetization is not expected to deviate from linearity up to very high fields $g \mu_B H \sim J$. As shown in the inset of fig.~\ref{fig_magnetization}, this kagome plane susceptibility is much lower than the macroscopic susceptibility at the same $T=1.7$~K but significantly higher than the local susceptibility obtained from $^{17}$O NMR. However if the
effective moment of the extended paramagnetic defects induced by spin vacancies in the kagome lattice is small, they may not be
saturated in the maximum applied field of 14~T and they then also contribute to the linear magnetization. As a result the extracted
susceptibility must be considered as an upper bound to the intrinsic kagome plane susceptibility.

Second, a saturated magnetization which can be isolated by subtraction of the high field linear contribution (discussed above) from the measured total magnetization (see red curve in Fig.~\ref{fig_magnetization}). This part of the magnetization resembles a Brillouin function for a spin 1/2 although the saturation is reached at a slightly higher field suggesting some residual antiferromagnetic interactions of $\sim 1$~K~\cite{BertHerbert,MendelsHerbert,deVries}.  This second contribution can be safely attributed to the interlayer Cu$^{2+}$ defects. From the value of the saturated magnetization one gets about $7\%$  of such defects out of the total number of Cu, in perfect agreement with the direct and indirect estimates from other studies.

\begin{figure}[t]
\begin{center}
\label{fig_magnetization}
\includegraphics[width=24pc]{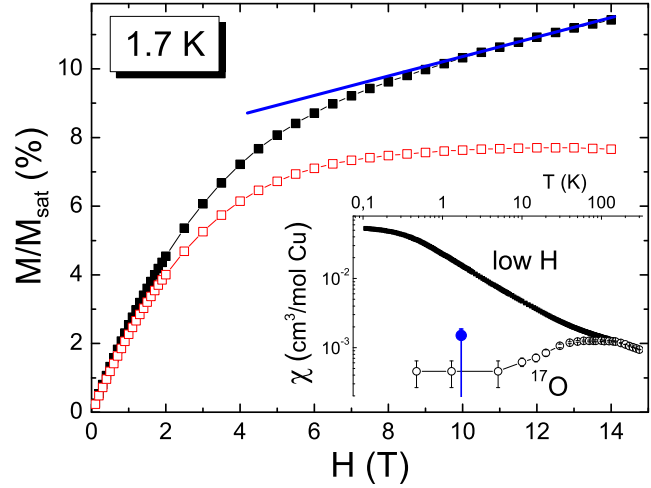}
\caption{\label{fig_magnetization} (Color on line) Field dependence of the 1.7~K
macroscopic magnetization (black squares) of Herbertsmithite
normalized to the saturated magnetization of 1 mole of Cu$^{2+}$
S=1/2 spins. The magnetization of the defective inter-layer
Cu$^{2+}$ spins (red open squares) is estimated by subtracting the
linear contribution of the kagome planes (blue line) from the total
magnetization. Inset : the kagome plane susceptibility at 1.7~K, i.e. the slope of the linear contribution, is
shown as the blue circle with large downward error bar on the plot
of the macroscopic susceptibility measured in 0.1~T (black curve) and $^{17}$O NMR
lineshift (open circles) reproduced from Fig.~\ref{fig_shift}. Adapted from ref.~\cite{BertHerbert}.}
\end{center}
\end{figure}

\subsubsection{Specific heat}

\begin{figure}[t]
\begin{center}
\includegraphics[width=22pc]{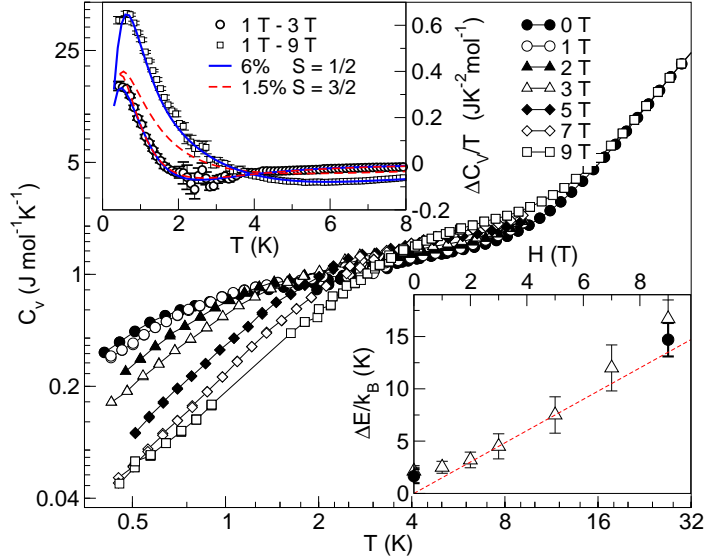}
\caption{(Color on line) The low temperature heat capacity of Herbertsmithite as a function of
applied field. Inset top: The temperature dependence of Cv(1 T) - Cv(3 T) and Cv(1 T) - Cv(9 T), fitted to two models,
$S=1/2$ and $S=3/2$. The latter cannot
be fitted simultaneously with the Cv(1 T) - Cv(3 T) and Cv(1 T) - Cv(9
T) curves. Even for $S = 1$ the fit is less good than for $S = 1/2$. Inset bottom: field dependence of the $S=1/2$ doublet. Adapted from ref~\cite{deVries}.}
\label{Cv}
\end{center}
\end{figure}

The specific heat of Herbertsmithite was measured down to 0.1~K under various applied fields up to 14~T~\cite{Helton, deVries}. Since no isostructural non-magnetic compound is available, reliable conclusions about the magnetic contribution cannot be extracted above 5-10~K because of the dominant contribution from phonons. The evolution of the $T$ dependence of the specific heat under an applied field can be fitted satisfactorily with a Schottky model of quasi-free spins, which value is quite accurately $S=1/2$, assuming an amount of 6~\% of the total number of Cu. Including interactions between the "quasi-free" Cu on the Zn site might indicate that the analysis developed in \cite{MisguichHerbert} could be corrected through the addition of extra terms. Whether such corrections to a pure Schottky analysis would correct the small 2~K value measured for $\Delta E$ (fig~.~\ref{Cv}) or whether the latter reveals some intrinsic energy scale in Herbertsmithite  is still an open issue. From the $0.66<x\leq 1.0$ wide  range of compositions in the paratacamite family where no ordering occurs, and the weakness of interplanar interactions that they generate, it seems safe to assess that the intersite Cu cannot influence dramatically the susceptibility of the kagome planes. Using such a scenario and assuming in addition that the kagome susceptibility is unaffected by spinless Zn defects, a very detailed analysis, using series expansions and Pade approximants for the pure kagome lattice~\cite{MisguichHerbert} show that both low-$T$ susceptibility and specific heat data in Herbertsmithite can be fairly reproduced with a few \% weakly interacting spins. There are some discrepancies, especially for the deduced entropy which is found smaller than that of the pure kagome lattice. This might point at the need to include the response of the spinless defects or other extra terms (see discussion section) in the interpretation of these results.

Overall, due to the dominant contribution from these free centers, it is quite hard to conclude about the analytic dependence of the specific heat which behavior depends strongly on the selected $T$-range.

\section{Discussion}

\begin{figure}[t]
\begin{center}
\includegraphics[width=22pc]{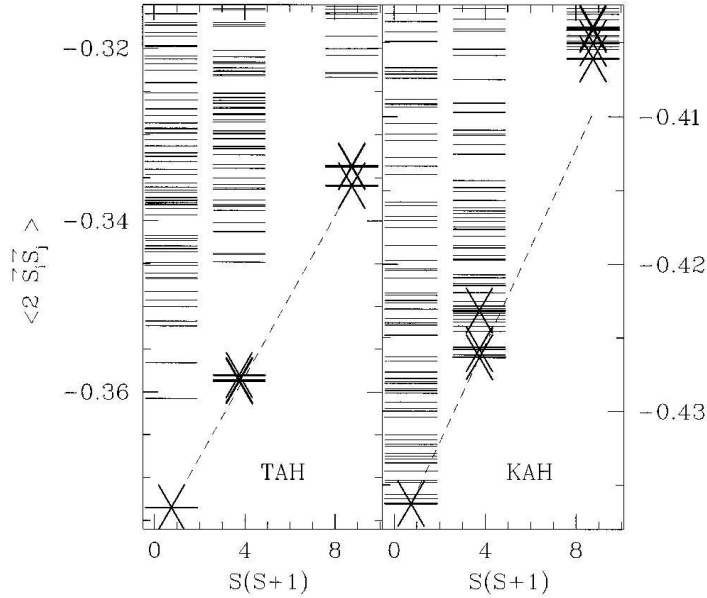}
\caption{Energy spectrum using exact diagonalizations techniques, from ref.~\cite{Lecheminant97}. The striking difference between the triangular and kagom\'e antiferromagnets, labeled TAH and KHAF, is the continuum of excitations in the singlet channel. The gap between the first singlet and triplet states is estimated at  $\Delta< J/20$, but might vanish at the thermodynamic limit~\cite{Lhuillier}.}
\label{ExcitationSpectra}
\end{center}
\end{figure}

\begin{figure}[t]
\begin{center}
\includegraphics[width=22pc]{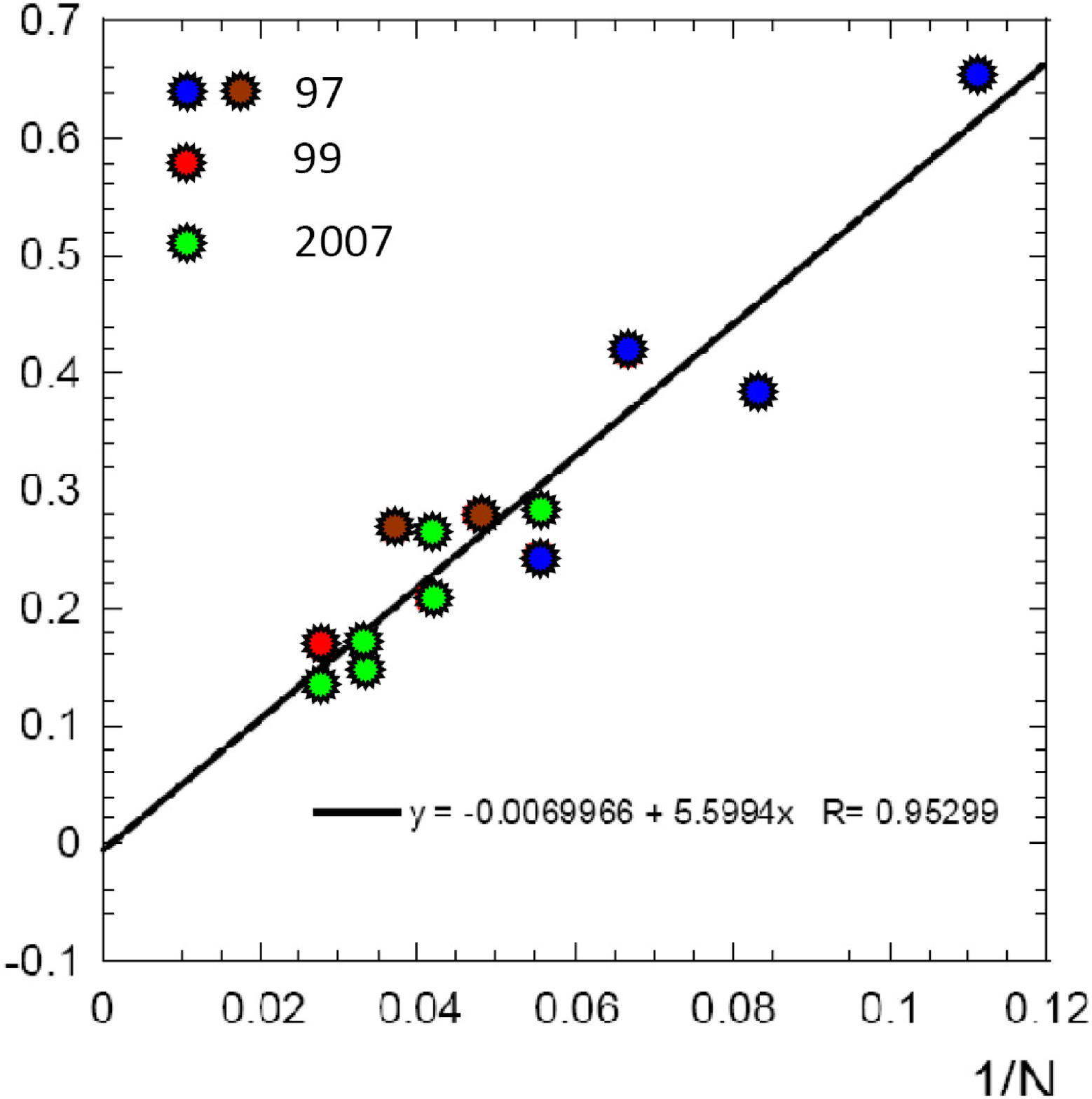}
\caption{(Color on line) Gap determined from exact diagonalizations techniques versus the inverse size of the sample. The sample size was doubled within 10 years. The gap might vanish at the thermodynamic limit~\cite{Lhuillier} as shown by the least square fit with a 3\% J accuracy. Unpublished data, courtesy from P. Sindzingre and C. Lhuillier.}
\label{GapNum}
\end{center}
\end{figure}

As emphasized in the introduction, the discovery of Herbertsmithite has triggered many new ideas about the ground state of quantum kagome antiferromagnets~\cite{Ran2007, Singh2007, Ryu}. Exact diagonalizations of finite clusters in the $S=1/2$ quantum limit had earlier suggested a picture of a fluctuating ground state with an unusual continuum of singlet excitations between a singlet ground state and the first excited triplet, at variance with the triangular HAF (Fig.\ref{ExcitationSpectra})~\cite{Lecheminant97, Waldtmann98} which features a more connected edge-sharing lattice. The central question of whether the small spin-gap $\lesssim J/20$ survives or vanishes at the thermodynamic limit is still pending~\cite{Lhuillier}, see fig.~\ref{GapNum}. A remarkable result is that an effective dimer-model yielded a similar energy landscape with the same entropy~\cite{Mila}. These are quite strong indications that rather than condensing into an ordered state, the system breaks into singlets which resonate to lower the total energy, possibly pointing at a RVB state. Quite recently, more proposals have been made for the ground state of the KHAF, but it is not the aim of this paper to review in detail the various models.

From an experimentalist point of view, extracting the existence and value of the gap is of crucial importance. The dynamical susceptibility which could be extracted from neutron scattering, its $Q$ dependence, the NMR relaxation rates and the specific heat are also stringent tests of theories although there are little predictions at the present stage about these quantities and their exact $T$-dependence.

Two main features are emerging from the experimental studies on Herbertsmithite, which are reported in the previous sections. The ground state possesses a finite magnetic susceptibility. Hence it is not a pure singlet state and, concomitantly, no spin-gap is observed in the excitations. One is therefore led to the central question of how near (or far) Herbertsmithite is from the ideal S=1/2 kagome Heisenberg antiferromagnet.

Various additional terms need to be incorporated into the Heisenberg Hamiltonian. We review them sequentially below but they likely have to be combined in order to account adequately for the present experimental data in Herbertsmithite. This might have a positive counterpart in the future since varying them through more advanced materials engineering might prove a rewarding way to isolate their influence on the intrinsic properties of the kagome lattice and could help discriminate between models.

\subsection{Applied field}
In the context where macroscopic susceptibility is dominated by the defects contribution at low $T$, the main measurement of the susceptibility relies upon local NMR measurements which are all performed in fields of the order of a few Teslas. This represents a few K energy for spins 1/2. The non-zero value of the susceptibility and the shape of its $T$-variation both indicate that in a scenario of a singlet-triplet gap, the value of the gap under an applied field, should be less than $J/200 \sim 1$~K. There is no calculation at present of how the field could decrease the value of the gap and this is an issue which should be addressed in the future.

\subsection{Dzyaloshinskii-Moriya (DM) anisotropy}

\begin{figure}[tb]
\begin{center}
\includegraphics[width=28pc]{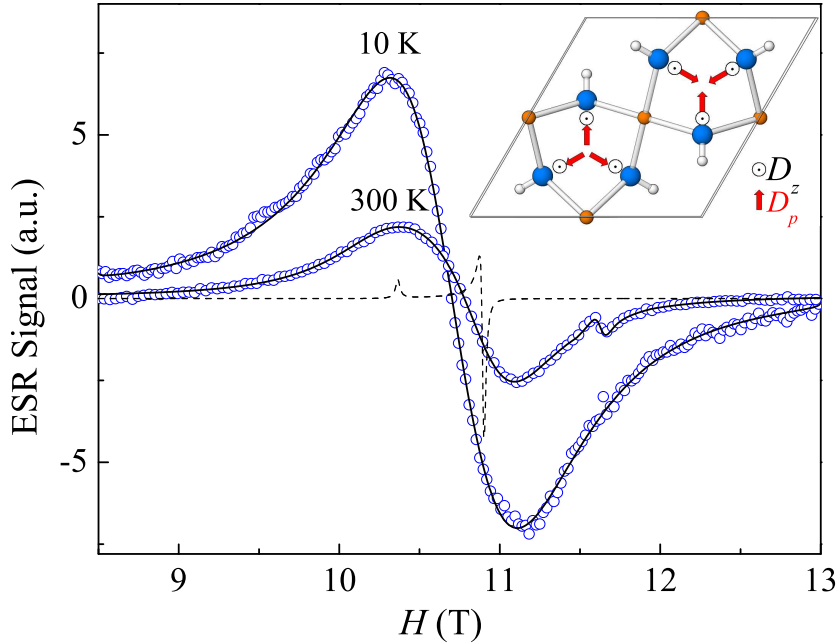}
\caption{(Color on line) ESR spectra taken at 300 and 10~K. The broad linewidth is due to a large anisotropy. The dashed line shows the anisotropy of the $g$- shift tensor, hence the minute deviation from a pure Heisenberg character of the spins, quite common for Cu$^{2+}$ ions. In the inset the Dzyaloshinskii vectors are represented. The Cu and O are respectively represented by small and large full circles (red and blue online). Adapted from ref.~\cite{Zorko2008}.}
\label{ESR1}
\end{center}
\end{figure}

\begin{figure}[tb]
\begin{center}
\includegraphics[width=28pc]{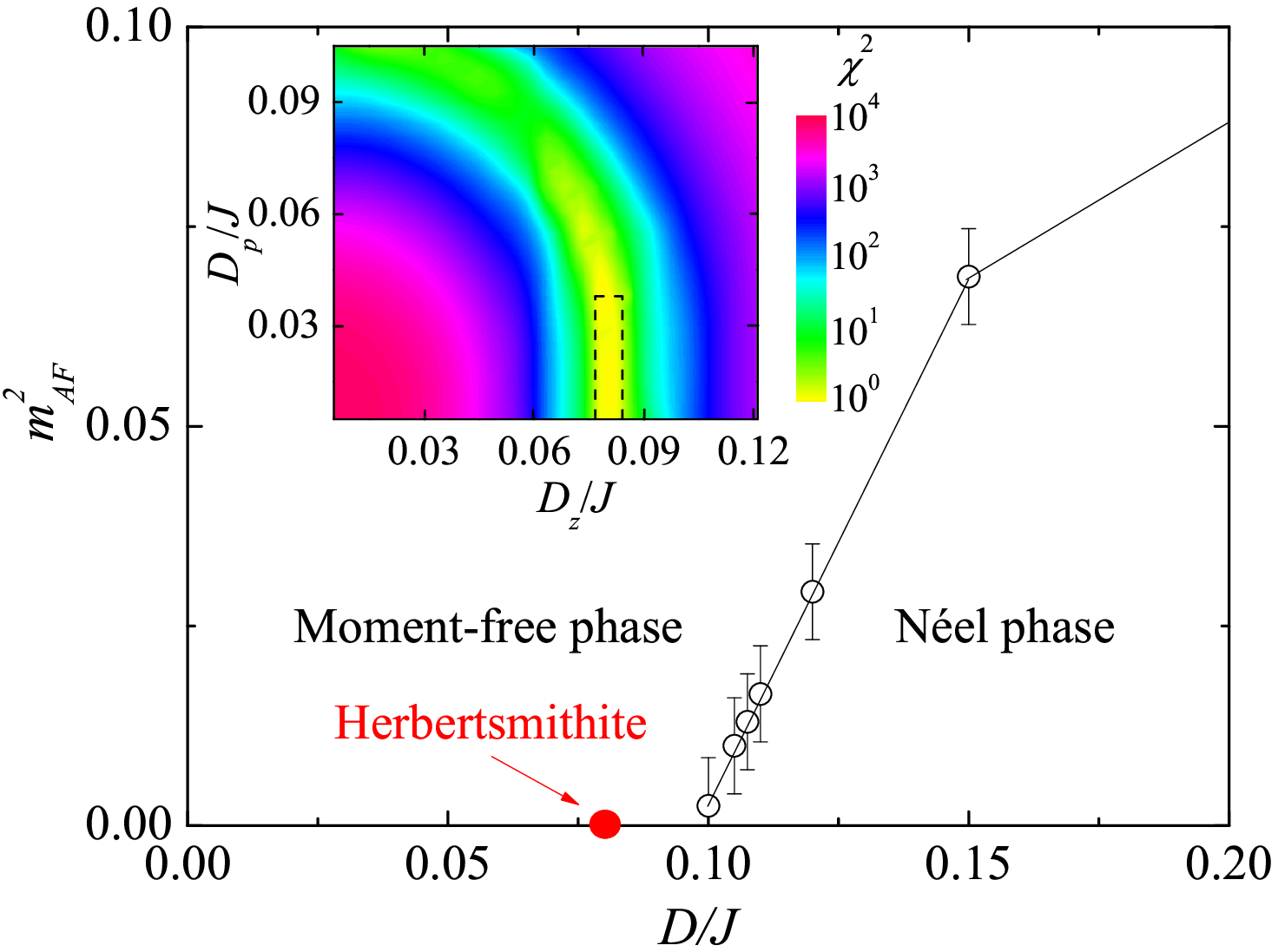}
\caption{(Color on line) Inset: Contour plot of the value obtained for $D_z$ and $D_p$ from the analysis of the high-$T$ ESR spectrum. Main panel: Phase diagram adapted from~\cite{Cepas}. The Herbertsmithite lies near the QCP where a N\'eel ordered phase would be observed. Adapted from ref.~\cite{Zorko2008}.}
\label{ESR2}
\end{center}
\end{figure}

Initially introduced in the context where the macroscopic susceptibility was assumed to be intrinsic, the DM anisotropy has been incorporated  into exact diagonalization calculations in order to account for the upturn of the susceptibility below $\simeq J$, a feature which is not expected in the unperturbed case~\cite{RigolSingh2}. The existence of a DM anisotropy $\overrightarrow{D} _{ij}.\overrightarrow{S}_i \times \overrightarrow{S}_j$ is granted since there is no inversion symmetry  between two adjacent Cu. $\overrightarrow{D}$ has two components, one along $z$, $D_z$, and one planar $D_p$ which act differently on the susceptibility and the specific heat (see Fig. ~\ref{Cv}). In order to explain the low-$T$ upturn of the susceptibility, one would need $D_p>|D_z|$, with $D_p \sim 0.3$~$J$. Yet, these values do not match those which were later extracted from the broad ESR line at room-$T$ (Fig.~\ref{ESR1}), {\it ie} a value of $D_p$ which does not exceed 0.05-0.1 $J$~\cite{Zorko2008} (Fig.~\ref{ESR2}).
Therefore only 15-30~\% of the upturn can be accounted for by DM anisotropy, which validates the defect approach. Within the latter approach, DM anisotropy induces a decrease of entropy which might solve one of the problems of the defect-only approach developed in ref.~\cite{MisguichHerbert}.

One of the most interesting outcome from the study of the DM anisotropy as a perturbation of the Heisenberg Hamiltonian is that an unexpected phase diagram appears Fig.~\ref{ESR2}). Whereas in the classical case, any minute amount of DM anisotropy will lead  to long-range order~\cite{Elhajal}, for spins $S=1/2$ a quantum critical point appears which is governed by the value of $D/J \sim 0.1$ where the system switches from a liquid to a N\'eel state \cite{Cepas, Rousochatzakis}. Noticeably, Herbertsmithite appears to be close to this quantum critical point.

The absence of a gap in the susceptibility is then well explained by a mixing of singlet and triplet states due to the DM interaction. Indeed with a value of the order of 0.1 $J$, DM interactions likely smear out a gap which is predicted to be $\lesssim J/20$.

\subsection{In-plane magnetic defects}

As developed in the previous section, some quasi-free Cu reside on the Zn site and dominate the low-$T$ macroscopic measurements, yet they do not fully account for the experimental observations. DM interactions can complete the view but spin vacancies in the kagome planes also generate a local magnetic response. This comes into play in the macroscopic measurements but yield specific signatures for local techniques. It is the aim of this paragraph to review shortly this topic with respect to the experimental results.

\begin{figure}[tb]
\begin{center}
\includegraphics[width=28pc]{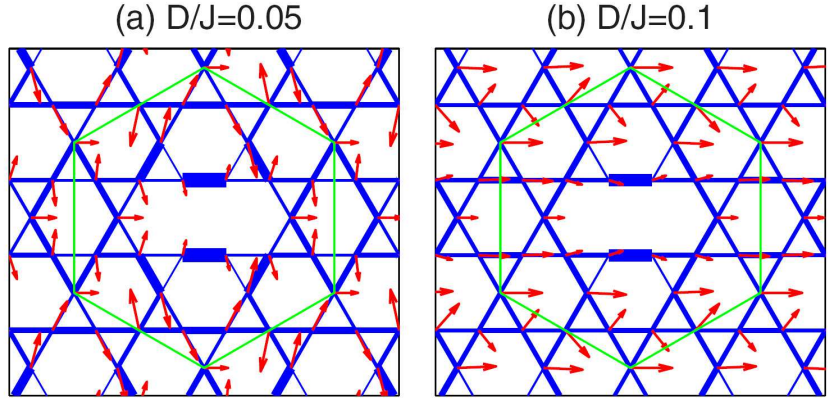}
\caption{(Color on line) Evolution of the in-plane magnetizations around a spin vacancy when $D$ increases. The calculations are performed under a field  $B=J/20$, similar to the applied field in NMR experiments and an angle of $30^{\circ}$ from the $z$ axis. The bond spin correlation values are represented by the thickness of the lines. The susceptibility remains small for the sites adjacent to a spin vacancy. Adapte from ref.~\cite{Rousochatzakis}. }
\label{figLausanne}
\end{center}
\end{figure}

Can the presence of 5~-~10\% dilution defects alter drastically the intrinsic physics? Down to 0.3~$J$, this does not seem to be much the case if one refers to numerical calculations of the macroscopic susceptibility and the specific heat~\cite{RigolSingh2}. More insights on the local and extended response around spinless defects come out from exact diagonalization studies of the effect of non-magnetic impurities in  kagome antiferromagnets, including the case where DM interactions are present. For $D=0$, a non-magnetic impurity is found to induce  a dimer freezing in the two adjacent triangles it belongs to, as expected from the relief of frustration~\cite{Dommange03}. Further from the impurity, a staggered response is generated like in all antiferromagnets~\cite{alloul07RMP}. The major difference with the unfrustrated case is that the response is not peaked on the sites next to the impurity. Both the tendency to dimer freezing and the staggered response hold for as large values of $D/J$ as 1, while the critical point at $D/J=0.1$ does not seem much affected by the presence of these spinless defects~\cite{Rousochatzakis} (Fig.~\ref{figLausanne}). Since for Herbertsmithite $D/J\lesssim 0.1$, one expects in $^{17}$O NMR a well defined susceptibility for the  oxygens next to a vacancy, reflected in the sharp "defect line" (D) whereas, combining the staggered response to the large number of defects, the susceptibility on further sites is expected to be more distributed as observed below 50~K on the main line (M). The experimental observations agree with that trend and the non-zero value of the $T\rightarrow 0$ susceptibility for the (D) line argue in favor of not being far from the quantum critical point, $0.06 <D/J<0.1$, in agreement with ESR analysis~\cite{Zorko2008,Rousochatzakis}.

\section{Conclusions}

Herbertsmithite is the only kagome compound not showing any magnetic transition  at least down to temperatures of the order of $10^{-4}J$. In this respect, it has opened wide the field of experimental investigations of the kagome physics, which has been at the center of theoretical researches on highly frustrated magnets for more than 20 years. Its ground state is not a simple singlet state and excitations are gapless. Considerable progress can be now anticipated in filling the gap between experiments and theory and overall in unveiling the ground state of the pure kagome antiferromagnet, especially through the investigation of the thermal dependence of the excitations. Models should take into account both the DM interaction ($\sim 0.1J$) which, due to low symmetry, might unfortunately not be avoided in any kagome based compounds and the existence of spinless defects ($\sim$ a few \%). Extensive exact diagonalization studies combining most of the experimental parameters have recently paved the way.  Working out the consequences of these perturbations might also help to understand and probe the kagome physics.

Certainly, the synthesis of single crystals and the controlled variation of both DM anisotropy and the amount of defects through new synthesis routes are highly desirable for further progress in the understanding of the ground state of the kagome antiferromagnet. We are still at the early era of a new story! More quantum frustrated compounds are being and will be synthesized.

\section{Acknowledgments}
We thank Pr. H.~Kawamura and Pr. S.~Maegawa for offering us the opportunity to present this review in this special issue.

We thank A. Olariu, M. de Vries and A. Zorko for their participation at various stages of the work performed by our group and for their help in the preparation of the manuscript.

We thank C. Lhuillier for discussions and a critical reading of our manuscript.

We deeply thank J. Helton, C. Lhuillier, F. Mila, I. Rousochatzakis, P. Sindzingre, M. de Vries and A. Zorko for providing original figures of their work for this review.

The work on Highly Frustrated Magnets at LPS has been supported by an ANR grant "OxyFonda" No. NT05-4-41913 and by the ESF research network "Highly Frustrated Magnetism". It has benefitted from the access to the Large Scale Muon Facilities PSI and ISIS, supported by the EC FP 6 program, Contract No. RII3-CT-2003-505925.


\begin{thebibliography}{99} 
\bibitem{Fazekas} P.~W.~Anderson, Mat. Res. Bull, {\bf 8}, 153 (1973), P.~Fazekas and P.~W.~Anderson, Philos. Mag. {\bf 30}, 423 (1974).
\bibitem{Anderson}  P.~W.~Anderson, Science, {\bf 235}, 1196 (1987).
\bibitem{Ramirez90} A.~P.~Ramirez, G.~P.~Espinosa and A.~S.~Cooper, Phys. Rev. Lett. {\bf 64}, 2070 (1990).
\bibitem{Uemura}Y.J.~Uemura {\it et al.}, Phys. Rev. Lett. {\bf 73}, 3306 (1994).
\bibitem{Levy} B. Levy, Physics Today, 16 (Feb. 2007).
\bibitem{Braithwaite} R~S~W Braithwaite , K.~Mereiter,~W.~Paar and A.~Clark A Mineral. Mag. {\bf 68} 527 (2004).
\bibitem{Shores} M.P.~Shores, E.A.~Nytko, B.M.~Barlett and D.G.~Nocera, J. A. Chem. Soc., {\bf 127}, 13462 (2005).
\bibitem{PLee} P.~A.~Lee, Science, Perspectives  {\bf 321}, 1306 (2008).
\bibitem{theory1}S.A.~Kivelson, D.S~Rokhsar and J.P.~Sethna, Phys. Rev. B {\bf 35}, 88865 (1987); R.~Moessner and S.L.~Sondhi, Phys. Rev. Lett. {\bf 86}, 1881 (2001)
\bibitem{theory2} M.B.~Hastings, Phys. Rev. B {\bf 63}, 014413 (2001); M. Hermele, T.~Senthil and M.P.A.~Fisher, Phys. Rev. B, {\bf 72}, 104404 (2005)
\bibitem{Zheng-b}X.G.~Zheng, H.~Kubozono, K.~Nishiyama, W.~Higemoto, T.~Kawae, A.~Koda and C.N.~Xu, Phys. Rev. Lett. {\bf 95}, 057201 (2005).
\bibitem{Zheng-a} X.G.~Zheng, T.~Kawae, Y.~Kashitani, C.S.~Li, N.~Tateiwa, K.~Takeda, H.~Yamada, C.N.Xu~and Y. Ren, Phys. Rev. B {\bf 71} 052409 (2005)
\bibitem{Kim}J.-H.~Kim, S.~Ji, S.-H.~Lee, B.~Lake, T.~Yildirim, H.~Nojiri, H.~ Kikuchi, K.~Habicht K, Y.~Qiu and K.~Kiefer, Phys. Rev. Lett. {\bf 101}, 107201 (2008).
\bibitem{clinoatacamite_Wills2}  A.S. Wills and J.Y.~Henry, J. Phys. Cond. Mat., {\bf 20}, 472206 (2008).
\bibitem{MendelsHerbert} P.~Mendels, F.~Bert, M.A.~de~Vries, A.~Olariu, A.~Harrison, F.~Duc, J.C.~Trombe, J.S.~Lord, A.~Amato and C.~Baines, Phys. Rev. Lett. {\bf 98}, 077204 (2007).
\bibitem{Helton} J.S.~Helton,K.~Matan, M.P.~Shores, E.A.~Nytko, B.M.~Bartlett, Y.~Yoshida, Y.~Takano, A.~Suslov, Y.~Qiu, J.H.~Chung, D.G.~Nocera and Y.S.~Lee, Phys. Rev. Lett. {\bf 98}, 107204 (2007).
\bibitem{Imai} T.~Imai, E.A.~Nytko, B.M.~Bartlett, M.P.~Shores and D.G.~Nocera, Phys. Rev. Lett. {\bf 100}, 077203 (2008).
\bibitem{Olariu} A.~Olariu, P.~Mendels, F.~Bert, F.~Duc F, J.C.~Trombe, M.~de~Vries and A.~Harrison, Phys. Rev. Lett. {\bf 100}, 087202 (2008).
\bibitem{Ofer} O.~Ofer, A.~Keren, E.A~Nytko, M.P.~Shores, B.M.~Bartlett, D.G.~Nocera, C.~Baines and A.~ Amato, Phys. Rev.B.{\bf 79}, 134424 (2009)
\bibitem{footnote} The weak superexchange constant $J'/J \sim 0.1$ was attributed to the proximity of the corresponding Cu-OH-Cu bond angle 97$^\circ$ to the antiferromagnetic/ferromagnetic limit according to Goodenough-Kanamori analysis~\cite{Shores}. In this context, one can expect a weak hyperfine coupling of $^{17}$O to the misplaced coppers in the triangular Zn planes.
\bibitem{RigolSingh1} M.~Rigol and R.~R.~P.~Singh, Phys. Rev. Lett.~{\bf 98}, 207204 (2007).
\bibitem{MisguichHerbert}G.~Misguich and P.~Sindzingre, Eur. Phys. J. B {\bf 59}, 305 (2007).
\bibitem{Mendels_SCGO} P.~Mendels {\it et al.}, Phys. Rev. Lett. {\bf 85}, 3496 (2000).
\bibitem{Limot} L.~Limot  {\it et al.}, Phys. Rev. B {\bf 65}, 144447 (2002).
\bibitem{Bono1} D.~Bono {\it et al.}, Phys. Rev. Lett. {\bf 92}, 217202 (2004).
\bibitem{Hiroi}Z.~Hiroi {\it et al.}, J. Phys. Soc Jpn {\bf 70}, 3377 (2001).
\bibitem{deVries} M.A.~de Vries, K.V.~Kamenev, W.A.~Kockelmann,J.~Sanchez-Benitez and A.~Harrison, Phys. Rev. Lett. {\bf 100}, 157205 (2008).
\bibitem{Lee07}S.-H.~Lee, H.~Kikuchi, Y.~Qiu, B.~Lake, Q.~Huang, K.~Habicht and K.~Kiefer, Nature Mater. {\bf 6}, 853 (2007).
\bibitem{Bert09} F.~Bert, A.~Olariu, A.~Zorko, P.~Mendels, J.C.~Trombe, F.~Duc, M.A. de Vries, A.~Harrisson, A.D.~Hillier, J.~Lord, A.~Amato and C.~Baines, J. of Physics {\bf 145}, 012004 (2009).
\bibitem{Chitra08RMN} M.J.~Rozenberg and R.~Chitra, Phys.Rev. B {\bf 78}, 132406 (2008).
\bibitem{alloul07RMP} H.~Alloul, J.~Bobroff, M.~Gabay and P.J.~Hirschfeld, Rev. Mod. Phys {\bf 81}, 45 (2009).
\bibitem{Bert04}F.~Bert, D.~Bono, P.~Mendels, J.C.~Trombe, P.~Millet, A.~Amato and C.~Baines, J. Phys.: Condens. Matter {\bf 16} S829--S834 (2004).
\bibitem{BertHerbert} F.~Bert, S.~Nakamae, F.~Ladieu, D.~L'Hote, P.~Bonville, F.~Duc, J.C.~Trombe and P.~ Mendels, Phys. Rev. B {\bf 76}, 132411 (2007).
\bibitem{Ran2007}Y.~Ran, M.~Hermele, P.A.~Lee~A and X.G.~Wen X~G, Phys. Rev. Lett. {\bf 98}, 117205 (2007).
\bibitem{Singh2007} R.R.P~Singh and D.A.~Huse, Phys. Rev. B {\bf 76}, 180407 (2007); ibid {\bf 77},144415 (2008).
\bibitem{Ryu} S.~Ryu, O.I.~Motrunich, J.~Alicea and M.P.A~Fisher, Phys. Rev. B {\bf75}, 184406 (2007).
\bibitem{Lecheminant97}P.~Lecheminant, B.~Bernu, C.~Lhuillier, L.~Pierre and P.~Sindzingre, Phys. Rev. B{\bf 56}, 2521 (1997).
\bibitem{Waldtmann98} C.~Waldtmann, H.U~Everts, B.~Bernu, C.~Lhuillier, P.~Sindzingre, P.~Lecheminant and  L.~Pierre, Eur. Phys. J. {\bf 2} 501--507 (1998).
\bibitem{Lhuillier} P.~Sindzingre and C.~Lhuillier, Eur. Phys. Lett. {\bf 88} 27009 (2009).
\bibitem{Mila}F.~Mila, Phys. Rev. Lett., {\bf 81}, 2356 (1998).
\bibitem{RigolSingh2}M.~Rigol and R.~R.~P.~Singh, Phys. Rev. B~{\bf 76}, 184403 (2007).
\bibitem{Zorko2008} A.~Zorko, S.~Nellutla, J.~van Tol, L.C.~Brunel, F.~Bert, F.~Duc, J.C.~Trombe, M.A.~de~Vries, A.~Harrison and P.~Mendels, Phys. Rev. Lett. {\bf 101}, 026405 (2008). A. Zorko \emph{et al} J. Phys.: Conf. Ser. {\bf 145}, 012014 (2009).
\bibitem{Elhajal} M.~Elhajal, ~B.~Canals and C.~Lacroix, Phys. Rev. B {\bf 66}, 014422 (2002).
\bibitem{Cepas}O.~C\'epas, C.~M.~Fong, P.~W.~Leung, C.~Lhuillier, Phys. Rev. B {\bf 78}, 140405 (2008).
\bibitem{Rousochatzakis}I.~Rousochatzakis, S.R.~Manmana, A.M.~Laeuchli, B.~Normand and F.~Mila, Phys. Rev. B {\bf 79}, 214415 (2009).
\bibitem{Dommange03} S.~Dommange, M.~Mambrini, B.~Normand and F.~Mila, Phys. Rev.B {\bf 68}, 224416 (2003).


\end{thebibliography}
\end{document}